\documentclass[aps,pre,amsmath,amssymb,groupedaddress,twocolumn,showpacs]{revtex4}
\usepackage{amssymb}
\usepackage{graphicx}

\begin{document}

\title{Arrested state of clay-water suspensions: gel or glass?}
\author{B.~Ruzicka$^{1}$, L.~Zulian$^{1}$, R. Angelini$^{1}$,
 M. Sztucki$^{2}$, A. Moussa\"id$^{2}$, G.~Ruocco$^{1}$}
\affiliation{ $^{1}$ SOFT INFM-CNR and Dipartimento di Fisica,
Sapienza
Universit$\grave{a}$ di Roma, I-00185, Italy.\\
$^{2}$ European Synchrotron Radiation Facility. B.P. 220 F-38043
Grenoble, Cedex France.
}
\date{\today}

\begin{abstract}
The aging of a charged colloidal system has been studied by Small
Angle X-rays Scattering, in the exchanged momentum range $Q$=0.03
$\div$ 5 nm$^{-1}$, and by Dynamic Light Scattering, at different
clay concentrations ($C_w=$0.6 \% $\div$ 2.8 \%). The static
structure factor, $S(Q)$, has been determined as a function of
both aging time and concentration. This is the first direct
experimental evidence of the existence and evolution with aging
time of two different arrested states in a single system simply
obtained only by changing its volume fraction: an inhomogeneous
state is reached at low concentrations, while a homogenous one is
found at high concentrations.

\end{abstract}
 \maketitle

In recent years, dynamical arrest in colloidal, and more generally
in soft matter systems, has gained increasing attention
\cite{ZaccarelliRev}. Specifically, much effort has been devoted
to clarify the dynamical behavior at large packing fraction, where
the dynamical arrest is commonly identified as a kind of "glass
transition". In colloids where both short range repulsion and
attraction are present, a rich phenomenology is found: a
re-entrant liquid-glass line, two kinds of glasses (named
"attractive" and "repulsive") and a glass-glass transition line
have been predicted and experimentally observed \cite{Colloids}.
Very recently an increasing attention has also been devoted to the
arrest at much smaller densities, in the usually called "gel
region" of the phase diagram \cite{ZaccarelliRev}. Experimental
and simulation \cite{PhaseSep} studies have proven that, for
hard-core plus spherically symmetric pair-wise attractive
potentials, arrest at low density occurs only through an
interrupted phase separation. More complex attraction is necessary
to produce {\it gelation in equilibrium} which takes place when
the gel state is reached continuously from an ergodic phase. This
occurs when a long-range repulsion, induced for example by
residual charges on the colloidal particles, is added to the short
range depletion attraction \cite{+LRrepulsion}. Very recently a
suppression of the phase separation has also been achieved by
using an anisotropic interaction potential
\cite{ZaccarelliJCP124}. In this scenario lowering the
"coordination number", i.e. the number of nearest neighbors
allowed by the interaction potential, pushes the spinodal line to
lower and lower packing fraction opening up the possibility to
reach very low temperature (and hence states with extremely long
bond lifetimes) without encountering phase separation. This
permits the formation of an {\it equilibrium bonding gel}, i.e. a
spanning network of long-living "physical" bonds at very low
colloids concentrations \cite{ZaccarelliRev}.

The study of different arrested states in colloidal systems is
therefore of crucial importance.  In fact gels and glasses have
often been viewed in an unifying framework due to unambiguous
similarities in the phenomenology accompanying the transition to
the kinetically arrested state. However, the limits of this
unifying scenario are emerging as the matter is further
investigated: a deeper comprehension of the differences and the
common features of gels and glasses in colloids is necessary and
lively debated in the up-to-date literature.

In Laponite suspensions -the system under investigation in the
present work- nanometric size disks form a charged colloidal
dispersion with a rich phase diagram. The competition between
attractive and repulsive interactions and/or the anisotropy of the
potential that originates the complexity of the phase diagram and
the existence of several aggregation processes. Therefore, in
recent years, Laponite suspensions have been widely studied not
only for the important industrial applications \cite{Laporte} but
especially for their peculiar experimental/theoretical properties
\cite{Mourchid, Kroon, Levitz,Mongondry1, Bonn, Knaebel,
Ruzicka,Bandyopadhyay,Li, Schosseler, Ianni,Joshi,Mossa}. The
first Dynamic Light Scattering (DLS) study \cite{Kroon} on this
system has shown a slowing down of its dynamics (aging) as it
evolves towards equilibrium. Therefore evolution with respect to
waiting time from a liquid to a gel/glass state has been usually
investigated for samples at clay concentration of $C_w=3 \%$ and
at salt concentration of $C_s=1 \times 10^{-4} M$. A recent
complete DLS study in a wide range of clay and salt concentrations
has permitted to individuate a surprising final arrested state not
only for high but also for very low clay concentrations
\cite{Ruzicka}, at variance to previous \cite{Mourchid} and recent
\cite{Bonn} proposed phase diagrams. The aging time evolution from
the initial liquid to the final arrested state occurs in a time
that strongly depends on salt and clay concentrations and that
increases as clay and/or salt concentrations are decreased
\cite{Ruzicka}, reaching the order of some months for $C_s=1
\times 10^{-4} M$ and $C_w=0.3 \%$. The differences in the aging
time evolution of both the raw spectra and the parameters obtained
by their analysis have also permitted to distinguish between two
different routes to reach two final non ergodic states at low and
high clay concentrations \cite{Ruzicka}(see Fig. 6 of
\cite{RuzickaPhil}). The mechanisms that originate the existence
of two different non ergodic states are a very interesting point.
Moreover, whether or not these two states correspond to really
different structures and what is the nature of these arrested
states are up to now open and intriguing questions.

In this letter we give the first direct experimental evidence of
the existence of two different arrested states in Laponite
suspensions and of their nature. The study of the aging time
evolution of both the dynamic and the static structure factors has
been performed through the combination of DLS and Small Angle
X-rays Scattering (SAXS) techniques. These measurements permit to
observe directly the evolution with aging time of the system and
to distinguish between final inhomogeneous and homogeneous states
resolving the longstanding controversy about the gel or glass
nature of Laponite arrested state \cite{Mongondry1,Bonn}. More
generally, while the signature for the time evolution of a gel was
quite well known \cite{Wilking} this is the first experimental
evidence of the $S(Q)$ time evolution for two different arrested
states obtained only by changing sample's volume fraction.

Laponite suspensions were prepared in a glove box under N$_{2}$
flux and were always kept in safe atmosphere to avoid samples
degradation \cite{Thompson}. The powder, manufactured by Laporte
Ltd, was firstly dried in an oven at $T$=400 C for 4 hours and it
was then dispersed in pure deionized water ($C_s\simeq 10^{-4}$
M), stirred vigorously for 30 minutes and filtered soon after
through 0.45 $\mu m$ pore size Millipore filters. The same
identical protocol has been strictly followed for the preparation
of each sample, fundamental condition to obtain reliable and
reproducible results as also recently reported by \cite{Cummins}.
The starting aging time (t$_w$=0) is defined as the time when the
suspension is filtered. Samples were placed and sealed in thin
glass capillaries with a diameter of 2 mm to be used both for DLS
and SAXS measurements. As already discussed, the waiting time
evolution of the aging dynamics requires few hours for high clay
concentrations and months for low ones. Therefore only the aging
of samples at high concentrations can be directly followed during
allocated SAXS beamtime. To investigate samples at low
concentrations, nominally identical concentration samples have
been prepared at different dates starting three months before the
planned experiment, so to have different waiting times at the
moment of the measurements.

DLS experiments were carried out using an ALV goniometer fitted
with a toluene bath. The incident laser wavelength was a diode
pumped, frequency doubled Nd-YAG 30 mW laser ($\lambda$ = 532 nm).
The scattered intensity was recorded with an avalanche photodiode.
The intensity correlation function was directly obtained as
$g_2(Q,t)=<I(Q,t)I(Q,0)>/<I(Q,0)>^{2}$, where $Q$ is the modulus
of the scattering wave vector defined as $Q=(4\pi
n/\lambda)sin(\theta/2)$. In the present experiment
$\theta$=90$^o$ and the acquisition time of each $g_2(Q, t)$ was
120 s.

SAXS measurements were performed at the High Brilliance beam line
(ID2) at the European Synchrotron Radiation Facility (ESRF) in
Grenoble, France using a 10 m pinhole SAXS instrument. The
incident x-ray energy was fixed at 12.6 keV. The form factor
$F(Q)$ was measured using a flow-through capillary cell. In SAXS
data analysis the corrections for empty cell and water have been
taken into account. The measured structure factor has been
obtained as $S^{M}(Q)=I(Q)/F(Q)$.
\begin{figure}[t]
\centering
\includegraphics[width=5.3cm, height=14cm]{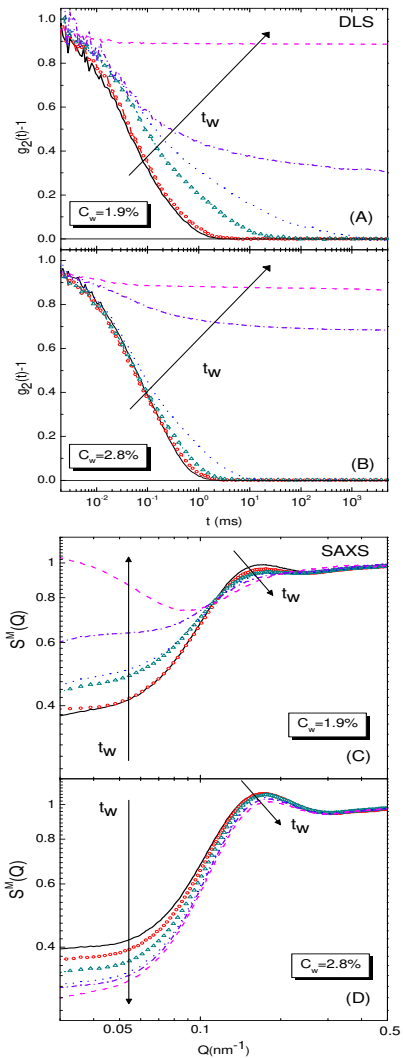}
\caption{(COLOR ONLINE) Evolution of the time autocorrelation
functions (top panels) and of the measured structure factors
(bottom panels) as a function of the waiting time $t_w$ (same
waiting times - same lines) for two different Laponite
concentrations: a low one, $C_w=1.9 \%$ ((A) and (C) panels) and a
high one, $C_w=2.8 \%$ ((B) and (D) panels).} \label{Fig1}
\end{figure}
Figure~\ref{Fig1} shows as an example the waiting time evolution
of both the clay density auto-correlation functions (top panels)
and the measured structure factors (low panels) for samples in the
low and high concentration regions, at $C_w=1.9 \%$ ((A) and (C)
panels) and $C_w= 2.8 \%$ ((B) and (D) panels) respectively. The
same sample was measured with the two techniques, DLS and SAXS
(same lines in the top and bottom panels correspond to roughly
same waiting time) and therefore a clear and direct comparison
between the dynamic and static structure factor behaviors can be
obtained. While, in fact, static light scattering measurements on
Laponite samples have been previously carried out
\cite{Kroon,Mongondry1, Li}, the full evolutions of the structure
factor with waiting times for both low and high concentration
samples have never been reported.

From Fig.~\ref{Fig1} the difference in the behavior of low and
high concentration samples is striking. For the high concentration
sample ((B) and (D) panels of Fig. ~\ref{Fig1}) there are no
significative changes in the structure factor profiles while the
sample crosses the ergodic/non ergodic transition (from dot to
dashdot and dash curves). As the system is performing aging there
is in fact only a progressive and slow decrease in the intensity
at very low $Q$ and a very small shift of the main peak to higher
$Q$ values. On the contrary, for the low concentration sample ((A)
and (C) panels of Fig.~\ref{Fig1}) there is an evident change of
the static structure factor as the sample ages and approaches the
non ergodic state (from dot to dashdot and dash curves)
specifically a progressive increased excess of scattering is
observed at low $Q$ values. Moreover, also a change in the shape
of the curve and a shift of the main peak to higher $Q$ values can
be recognized. These differences are the clear evidence that the
two samples are reaching the final non ergodic states following
different routes \cite{Ruzicka} and, furthermore, that the two
final arrested states are actually different.

\begin{figure}[t]
\centering
\includegraphics[width=.4\textwidth]{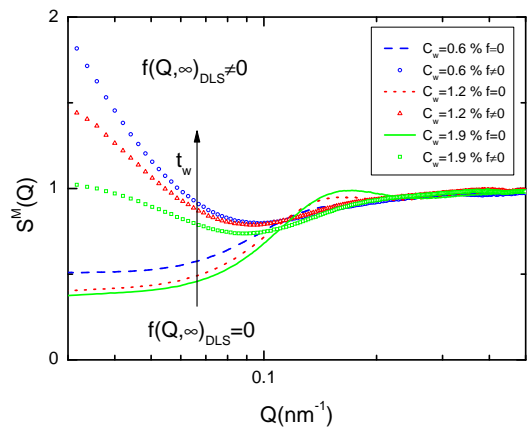}
\caption{(COLOR ONLINE)Waiting time evolution of the static
structure factors from full decay ($f(Q, \infty)=0$) to incomplete
decay ($f(Q, \infty)\neq 0$)in the corresponding DLS spectra for
three different low Laponite concentrations at $t_w\approx 1.2\\
t_w^{\infty}$.} \label{Fig2}
\end{figure}

The same behavior shown in panels A and C has been found for all
the investigated low concentration samples. In Fig.~\ref{Fig2} the
static structure factors measured for three different low
($C_w$$<$ 2.0 \%) Laponite concentrations in correspondence of the
full decay (small waiting time, lines) and incomplete decay (long
waiting time, symbols) of the DLS spectra are shown. Due to
different aging "velocity", we show here the $S^M(Q)$ at the same
$t_w/t_w^\infty$, being $t_w^\infty$ the $C_w$-dependent arrest
time, as defined in Ref. \cite{Ruzicka}. It is clear that in the
low $Q$ region the same increase of intensity is found as the
sample arrests. Moreover this excess of scattering at low $Q$
increases as Laponite concentration is decreased. This indicates
the existence of strong inhomogeneity in the suspension, that we
attribute to the formation of a network which -as the aging time
goes by- grows and eventually forms a gelled network. Also
observable is the disappearance of the main peak and/or its shift
to higher $Q$ values. On the contrary, none of these features is
observed for the high concentration  ($C_w$$>$ 2.0 \%) samples
where all the measurements have shown the same trend as the one
drawn in panel (D) of Fig.~\ref{Fig1}. This indicates the
homogeneity of the high $C_w$ arrested state. The observed
concentration dependence of the low $Q$ scattering intensity seems
to suggest that the transition between the two different final
states is not discontinuous but the arrested state becomes more
and more homogeneous increasing clay concentration.
Summarizing, the data reported here give a clear proof of the
existence of two structurally different arrested states in
Laponite suspensions, completing the information -already reported
in literature \cite{Ruzicka}- that low and high clay
concentrations reaches the arrested state following two different
routes. Static structure factor measurements, indeed, show an
excess of scattering in the low $Q$ region for the low
concentration samples, thus indicating the formation of an
inhomogeneous state, while for the high concentration samples the
$S^M(Q)$ data are in agreement with the formation of a homogeneous
arrested state.

The present experimental findings, together with the phenomenology
associated to gel and glass formation in colloids allow to
speculate on the nature of the observed arrested states. Despite
the apparent simple definition that "colloidal gels are arrested
state of matter at low density where the particles are tightly
bonded to each other so that the thermo-reversible bonds is
comparable or longer than the experimental observation time", the
essence of the (colloidal) gel state is still under debate
\cite{SciortinoStatPhys}. In this context, it  is important to
stress that sometimes the discussion on what is a gel and a glass
becomes a nominalistic dispute. In fact, the formation of a
bonding gel, which in a part of the colloidal community provides
indication of gel formation, for another part of the scientific
community, would be rather classified as glass formation
\cite{SciortinoStatPhys}. In this sense we can follow the
proposition that in a "gel" the attraction between colloidal
particles is the leading mechanism for gelation, while a "glassy
state" can be driven either by repulsion (hard-sphere or Wigner
glass) or by attraction (attractive glass). Moreover, while the
gel state is characterized by structural inhomogeneities
(signalled by a non-trivial low-$Q$ signal in the scattering
intensity) the glass is structurally homogeneous
\cite{ZaccarelliRev}.

On the ground of the previous discussion the low clay
concentration inhomogeneous arrested state can be considered as a
gel while the high concentrated homogeneous as a glass. Moreover,
in Laponite, increasing salt concentrations the arrested states
are still observed and occur even more rapidly \cite{Mongondry1,
Ruzicka}. This observation is in clear contradiction with the
formation of a colloidal glass induced by strong repulsive
interactions that are decreased (due to effect of screening) as
the ionic strength is increased. Therefore, the attractive
interaction must be responsible of the final arrested state. For
these reasons we can call the homogeneous state at high clay
concentrations an attractive glass, i.e. a homogeneous state
governed by attractive interactions (that can also be called
homogeneous gel). On the contrary, in the low concentration
region, the inhomogeneous state is compatible with the formation
of a gel state where the aging dynamics is supposed to be
characterized by the growth of region of strongly correlated
platelets which increases in size and, eventually, span all over
the systems at the gel transition.

It is worth to note that the features of this concentration region
fit with those of the newly discovered {\it equilibrium bonding
gel} region \cite{ZaccarelliRev}, a state found in low density
colloids with anisotropic interactions. In fact, the interaction
potential between Laponite platelets is certainly non spherical,
with privileged relative orientation, an aspect that probably has
not been properly accounted for in previous studies. Moreover, the
waiting time evolution of $S(Q)$ in Fig.~\ref{Fig1}C strongly
resembles the temperature evolution of $S(Q)$ of Fig.5 (a) of
Ref.\cite{ZaccarelliJCP124} obtained in simulation of a
valence-limited colloidal system. The increased intensity at small
wave vectors of $S(Q)$ as waiting time (temperature) is increased
(decreased) indicates that the system becomes more and more
compressible, with large inhomogeneities that can be seen as an
echo of the nearby phase separation or, equivalently, as a
consequence of building up a fully connected network. In fact one
can suppose that as the time goes by the system feels more and
more the attraction approaching, through successive equilibrium
states, the final arrested state as the ideal systems
\cite{ZaccarelliJCP124} do decreasing temperature. Also the
evolution of the $S(Q)$ with concentration seems to be in
qualitative agreement with Fig.5(b) of
Ref.\cite{ZaccarelliJCP124}: moving away from the spinodal line
(increasing $C_w$) the $Q\rightarrow 0$ peak decreases and the
nearest neighbor peak grows, signaling the increasing importance
of the packing. It is important to underline that, besides the
qualitative agreement between the behavior of the model and
Laponite suspensions no quantitative comparison can be done
between the systems due to Laponite peculiarity. Therefore, at
present, we are not also able to locate the spinodal line in
Laponite suspensions. It could be in fact possible that some of
the samples at very low concentrations are "inside" the spinodal
line and that for those systems a separation phase at very long
waiting times could take place.

In conclusion, the results reported here, resolve the longstanding
controversy about the final arrested state in Laponite suspensions
\cite{Mongondry1, Bonn}: both an inhomogeneous and a homogeneous
state exist. They are reached -as the system ages- following two
different dynamic routes \cite{Ruzicka}. The existence of two
different arrested states in this charged colloidal system with a
non spherical potential (further complicated by the presence of
charges), depending only on its concentration, gives a proof of
the strong connection existing between the "gel" and "glass"
states. A common experimental and theoretical effort for the
understanding of the still puzzling liquid-gel/glass transition
and a theoretical investigation about the features of the
different arrested states in presence of charged non spherical
potentials, are needed.

We acknowledge P. Bartlett, T. Nicolai, F. Sciortino and E.
Zaccarelli for helpful discussions.

\end{document}